\pdfoutput=1

\documentclass[11pt]{article}
\RequirePackage[OT1]{fontenc}
\RequirePackage{amsthm,amsmath}
\RequirePackage[numbers]{natbib}
\RequirePackage[colorlinks,citecolor=blue,urlcolor=blue]{hyperref}

\usepackage[utf8]{inputenc} 
\usepackage[T1]{fontenc}    
\usepackage[colorlinks]{hyperref} 
\usepackage{url}            
\usepackage{booktabs}       
\usepackage{amsfonts}       
\usepackage{amsmath}
\usepackage{nicefrac}       
\usepackage{microtype}      
\usepackage{color}
\usepackage{bbm}
\usepackage{amssymb, enumerate}
\usepackage{makecell}
\usepackage{wrapfig}
\usepackage{extarrows}
\usepackage{arydshln}

\usepackage{caption}
\usepackage{multirow}

\usepackage{amsmath,amsfonts,amsthm, amssymb, enumerate}
\usepackage{lipsum}
\usepackage{caption}
\usepackage{amssymb,graphicx,subfigure}
\usepackage[bottom]{footmisc}
\usepackage[dvipsnames]{xcolor}
\usepackage{array,multirow}
\usepackage{enumitem}

\usepackage{microtype}
\usepackage{graphicx}
\usepackage{subfigure}
\usepackage{booktabs} 
\usepackage{kotex}
\usepackage{graphicx}
\usepackage{amsmath}
\usepackage{amssymb}
\usepackage{booktabs}  
\usepackage{times}
\usepackage{epsfig}
\usepackage{graphicx}
\usepackage{amsfonts, amsthm, enumerate} 
\usepackage[bottom]{footmisc} 
\usepackage{enumitem}
\usepackage{setspace}
\usepackage{wrapfig}
\usepackage{extarrows}
\usepackage{makecell}
\usepackage{bbm}  
\usepackage{textcomp}
\usepackage{xcolor} 
\usepackage{caption}

\usepackage[font=small,labelfont=bf]{caption}

\usepackage{algorithm,algpseudocode}
\algnewcommand\TR{\item[{\textbf{Training phase}}]}
\algnewcommand\TE{\item[{\textbf{Test phase}}]}
\algnewcommand\Input{\item[{{Input:}}]}
\algnewcommand\Output{\item[{{Output:}}]}
\algnewcommand\Initialize{\item[{{Initialize:}}]}
\algnewcommand{\return}[1]{
	\State \textbf{return:}
	\Statex \hspace*{\algorithmicindent}\parbox[t]{.8\linewidth}{\raggedright #1}
}



\usepackage{setspace}
\usepackage[left=2cm,right=2cm,top=4cm,bottom=4cm,a4paper]{geometry}

\begin{document}
	\title{Enhanced artificial intelligence-based diagnosis using CBCT with internal denoising: Clinical validation for discrimination of fungal ball, sinusitis, and normal cases in the maxillary sinus}
	\date{}
	\author{\parbox{\linewidth}{\centering
		Kyungsu Kim$^{1,2}$\thanks{These authors contributed equally to this work as co-first authors.}, \, Chae Yeon Lim$^{3}$\footnotemark[1], \, Joongbo Shin$^{4}$, Myung Jin Chung$^{1,2,3,5}$, \, Yong Gi Jung$^{1,2,4}$\thanks{Corresponding author (ent.jyg@gmail.com)}\\ 
		{\small $^{1}$Medical AI Research Center, Research Institute for Future Medicine, Samsung Medical Center, Seoul, Republic of Korea}\\
		{\small $^{2}$Department of Data Convergence and Future Medicine, Sungkyunkwan University School of Medicine, Seoul, Republic of Korea}\\
		{\small $^{3}$Department of Medical Device Management and Research, SAIHST, Sungkyunkwan University, Seoul, Republic of Korea}\\ 
		{\small $^{4}$Department of Otorhinolaryngology-Head and Neck Surgery, Samsung Medical Center, Sungkyunkwan University School of Medicine, Seoul, Republic of Korea} \\
		{\small $^{5}$Department of Radiology, Samsung Medical Center, Sungkyunkwan University School of Medicine, Seoul, Republic of Korea} 
	}}

	\maketitle
\begin{abstract}

\noindent\textbf{Background and objective} $\,\,$ The cone-beam computed tomography (CBCT) provides three-dimensional volumetric imaging of a target with low radiation dose and cost compared with conventional computed tomography, and it is widely used in the detection of paranasal sinus disease. However, it lacks the sensitivity to detect soft tissue lesions owing to reconstruction constraints. Consequently, only physicians with expertise in CBCT reading can distinguish between inherent artifacts or noise and diseases, restricting the use of this imaging modality. The development of artificial intelligence (AI)-based computer-aided diagnosis methods for CBCT to overcome the shortage of experienced physicians has attracted substantial attention. However, advanced AI-based diagnosis addressing intrinsic noise in CBCT has not been devised, discouraging the practical use of AI solutions for CBCT. We introduce the development of AI-based computer-aided diagnosis for CBCT considering the intrinsic imaging noise and evaluate its efficacy and implications.

\noindent\textbf{Methods} $\,\,$ We propose an AI-based computer-aided diagnosis method using CBCT with a denoising module. This module is implemented before diagnosis to reconstruct the internal ground-truth full-dose scan corresponding to an input CBCT image and thereby improve the diagnostic performance. The proposed method is model agnostic and compatible with various existing and future AI-based denoising or diagnosis models. 

\noindent\textbf{Results} $\,\,$  The external validation results for the unified diagnosis of sinus fungal ball, chronic rhinosinusitis, and normal cases show that the proposed method improves the micro-, macro-average area under the curve, and accuracy by $7.4$, $5.6$, and $9.6\%$ (from 86.2, 87.0, and 73.4 to 93.6, 92.6, and 83.0\%), respectively, compared with a baseline while improving human diagnosis accuracy by 11\% (from 71.7 to 83.0\%), demonstrating technical differentiation and clinical effectiveness. In addition, the physician's ability to evaluate the AI-derived diagnosis results may be enhanced compared with existing solutions.

\noindent\textbf{Conclusion} $\,\,$ This pioneering study on AI-based diagnosis using CBCT indicates that denoising can improve diagnostic performance and reader interpretability in images from the sinonasal area, thereby providing a new approach and direction to radiographic image reconstruction regarding the development of AI-based diagnostic solutions. Furthermore, we believe that the performance enhancement will expedite the adoption of automated diagnostic solutions using CBCT, especially in locations with a shortage of skilled clinicians and limited access to high-dose scanning.
\end{abstract}

\section{Introduction}

\subsection{Clinical value of cone-beam computed tomography as an alternative to conventional computed tomography}
Chronic rhinosinusitis (CRS) \cite{benninger2003adult} is an inflammatory condition that affects from 2\% to 16\% of the population in the United States, being a significant disease that costs approximately \$10 billion in social expenses annually \cite{caulley2015direct}. To reduce the social burden caused by CRS, accurate early diagnosis is crucial through methods such as sinonasal computed tomography (CT), which also allows the establishment of a surgical plan for endoscopic sinus surgery \cite{cashman2011computed}. Conventional multidetector helical CT (MDCT) provides accurate high-quality coronal and sagittal images of the sinus \cite{bisdas2004three}, being a common diagnostic modality. However, MDCT has several drawbacks, including high cost, large installation area, and high radiation exposure \cite{albert2013radiation}. In addition, MDCT is time-consuming when capturing and reassembling pictures, possibly causing adverse reactions such as claustrophobia in patients. Radiation exposure to radiation-sensitive organs, such as the eyes and thyroid gland, cannot be prevented during MDCT, restricting its repeated prescription around those organs. 

Cone-beam CT (CBCT) is attracting attention as an image restoration technique that can overcome the shortcomings of MDCT. Compared with MDCT, CBCT features a shorter imaging time, less radiation dose, simpler reexamination, lower price, reduced installation area, and a simpler procedure even for claustrophobic patients owing to its open design. As CBCT employs an isotropic voxel to emit cone-shaped X-rays, they are irradiated simultaneously and not sequentially.
Therefore, CBCT can reduce the dose with fewer scans while accurately reconstructing the patient’s three-dimensional (3D) structures; that is, it retains the ratio between axes of the 3D volume from a target. In contrast, MDCT employs a linear X-ray beam and thus requires extensive imaging, resulting in large-dose exposure. The average dose used for MDCT imaging is 42\% higher than that used for CBCT imaging (108 vs. 63 μSy) \cite{de2015comparative}. Owing to its expansive potential area, CBCT can reconstruct an MDCT image from an irradiated region with a single scan, thereby achieving low-dose, high-speed, and high-resolution imaging \cite{yoo2006dosimetric,farman2009basics}.  

\subsection{Limitation of CBCT for widespread adoption in clinical use: Unreliable diagnosis in non-skeletal regions including soft tissues} 

As CBCT uses cone X-ray beams rather than linear beams, it can provide the aforementioned advantages. Therefore, it replaces conventional MDCT and panoramic radiography, especially in dental practice. However, the CBCT principle contradicts the Fourier slice theorem from a technical perspective \cite{makins2014artifacts}. Filtered back projection \cite{feldkamp1984practical}  is the gold standard reconstruction method for MDCT images intended to address the inverse problem based on plane projection instead of cone-beam projection. Most commercially available image restoration systems for CBCT use filtered back-projection because no solution is available to completely solve the inversion of cone-beam projections completely. Unlike MDCT, CBCT has intrinsic artifacts \cite{naitoh2013metal} and noise because it reconstructs an image that is close to the ground truth but cannot reconstruct the true image. The imaging of soft tissues is particularly susceptible to these technical limitations. Combined with small radiation doses, CBCT provides inferior image resolution and distortion compared with MDCT when imaging soft tissues with lesions \cite{zoumalan2009flat}. This limits the ability of CBCT to accurately diagnose inflammation in these tissues \cite{agrawal2012new}.

Various studies have claimed the need for educating clinicians to distinguish between disease and noise owing to the notable artifacts and noise on CBCT images. As the problem of low-quality soft tissue imaging with CBCT is almost nonexistent when imaging non-soft tissues, CBCT has been widely used in dentistry, where bone tissue is predominantly observed. \cite{alamri2012applications} In contrast, the low quality of CBCT for imaging soft tissues hinders the identification of lesions \cite{almeida2011soft} in medical specialties such as otorhinolaryngology. Consequently, the utility of CBCT has been limited in those specialties despite its advantages. In otorhinolaryngology, for instance, surgery is required for disease control of the sinonasal fungal ball because it is a frequent disease in the maxillary sinus that does not disappear spontaneously. Hence, simple CRS should be accurately distinguished from fungal balls in the early stage of the disease, and CT is essential for such diagnosis \cite{pagella2007paranasal}. However, CBCT is rarely used because of its insensitivity to soft-tissue differentiation in the maxillary sinus. Reliably detecting lesions in soft tissues using CBCT remains challenging and requires assessment by experienced specialists who have observed and treated several patients with the same lesion type by comparing paired CBCT and MDCT images.

\subsection{Artificial intelligence to overcome shortcomings of CBCT} 

In recent years, numerous studies on artificial intelligence (AI)-based automatic lesion diagnosis in recent years have demonstrated high diagnostic performance comparable to that of an experienced physician \cite{ahuja2019impact} or diagnostic support to physicians by AI solutions. Accordingly, AI has been applied to CBCT, achieving high quality and useful automatic diagnosis \cite{serindere2022evaluation, hiraiwa2019deep, setzer2020artificial, brignardello2020artificial, yang2022development}. AI-based processing can be viewed as automatically separating noise and disease information in a CBCT image to then identify the disease type by focusing on the corresponding information. As lesions and noise in CBCT images show distinctive patterns, AI-based computer-aided diagnosis (CAD) may help to identify a lesion using CBCT more accurately or effectively than a physician. Thus, AI-based CAD may effectively compensate for noise and artifacts in CBCT processing given the ability of AI to separate noise from the target lesion and provide more accurate diagnostic information to physicians than manual CBCT analysis.

Despite the technological aspirations, available AI-based CAD technology for CBCT has only been successfully applied to diseases related to the skeletal system, which can be imaged even by CBCT. However, no studies have achieved superior diagnosis of soft tissue lesions using CBCT, for which unmet demands in clinical fields beyond dentistry persist \cite{hiraiwa2019deep, setzer2020artificial, brignardello2020artificial, yang2022development}. Serindere et al. \cite{serindere2022evaluation} evaluated the diagnostic performance of CBCT by applying AI-based CAD to maxillary sinusitis, a lesion that primarily affects soft tissues. Although they confirmed the superiority of AI-based CAD using CBCT in comparison with panoramic images, no absolute advantage or technical difference between AI-based CAD using CBCT and other imaging modalities has been verified.

\subsection{Our contributions}
AI-based CAD using CBCT can reduce the inherent CBCT noise and then identify a disease or other health conditions in a noise-free environment. However, such an approach has not been devised, impeding the confirmation of its superiority and efficacy. Except for the aforementioned studies, to the best of our knowledge, AI solutions for CBCT have not been focused on diagnosis but instead included image restoration \cite{hwang2020very, wurfl2018deep, eulig2021deep, shen2022geometry, thies2020learning, shen2019patient, hauptmann2020multi, liang2020use} and segmentation  \cite{lahoud2021artificial, lin2021micro, sherwood2021deep, morgan2022convolutional}. Overall, no study has demonstrated the clinical utility of the advancement of AI-based CAD methods using CBCT. Nevertheless, it seems possible to diagnose soft tissue lesions to outperform existing AI-based CAD using CBCT. 

The contributions of this study to meet technical and clinical requirements for AI-based CAD using CBCT are summarized as follows: 
\begin{itemize}[noitemsep,topsep=0pt,leftmargin=3.5mm]
    \item Unlike existing developments \cite{ezhov2021clinically,lee2020diagnosis,du2022combined,lee2020automated}, which perform CAD by merely applying existing AI technology to CBCT images, our AI-based CAD method using CBCT incorporates an AI-based denoising stage.  Furthermore, existing denosing technique studies on CBCT \cite{KARIMI201671,CHAO2022536,liu2020cbct} do not apply them to the diagnostic technology, whereas our study is the first for this application, demonstrating its technical distinctiveness and clinical superiority.
    \item We apply and evaluate the proposed method to discriminate fungal balls, sinusitis, and normal cases in the maxillary sinus. Fungal balls in the maxillary sinuses constitute a typical condition whose early diagnosis is crucial considering its bad prognosis, especially for the elderly \cite{ferguson2000fungus}. Condition scanning using CBCT has attracted attention owing to the reduced radiation risk to the patient. However, existing methods for lesion localization on CBCT images in the presence of noise may be inaccurate, and physicians cannot rely on the CAD findings. In contrast, the proposed AI-based CAD system synthesizes a denoised full-dose reference MDCT image from the corresponding CBCT scan using an embedded denoising method to then accurately localize lesions. The proposed method may facilitate and increase the accuracy of diagnosis, supporting physicians in understanding automatic diagnosis data.
    \item We demonstrate that the proposed AI-based CAD method for CBCT achieves superior diagnostic performance than an existing method, with improvements in the micro-average area under the curve (AUC), macro-average AUC, and accuracy by $7.4$, $5.6$, and $9.6\%$ (from 86.2, 87.0, and 73.4 to 93.6, 92.6, and 83.0\%), respectively. In a comparative evaluation with assessments by six otorhinolaryngology resident physicians, the proposed method showed improved accuracy by 11.3\% (from 71.7 to 83.0\%), demonstrating its clinical value. For a fair evaluation, we used data from Samsung Seoul Medical Center for training (i.e., internal set) and Samsung Changwon Medical Center for evaluation (i.e., external set). 
    \item We demonstrate that using medical imaging equipment capable of concurrently performing CBCT and MDCT image restoration has great prospective applications. For instance, training AI-based CAD based on pairs of collected CBCT and MDCT images may allow to subsequently perform efficient diagnosis solely using CBCT (with a low dose) while omitting both MDCT imaging and additional training. Our development and findings demonstrate the feasibility of this approach.
\end{itemize} 
 
\section{Methods}
\subsection{Data acquisition and preprocessing}
\subsubsection{Data acquisition and annotation}
\label{sec:data_ac_an}
\begin{table}[ht] 
\vskip -4pt 
\caption{Number of 3D stacks for OMU R-MDCT per class in internal and external sets}
\footnotesize
\centering
{
	\resizebox{0.5\linewidth}{!}{
		\begin{tabular}{c|c|c}
            \hline
			\bf{Data type} & \bf{Internal set (\emph{n})} & \bf External set (\emph{n})     \\
            \hline
			HC & $130$ & $20$\\
            \hline
			CRS & $128$ & $18$\\
            \hline
            MFB & $254$ & $26$\\
            \hline
			Total & $512$ & $64$\\
            \hline
		\end{tabular}
  }
}
\label{Materials}
\end{table}

Real MDCT (R-MDCT) data were collected to synthesize pseudo-CBCT (P-CBCT) data. R-MDCT data consisted of internal and external sets, which were collected from Samsung Seoul Medical Center (Seoul, Republic of Korea) and Samsung Changwon Medical Center (Changwon, Republic of Korea), respectively, after approval by the Institutional Review Board (approval number: 2020-07-173). At 120 kVP and 2 mm-thick tubular surface slices, R-MDCT images from each institution were acquired using the same parameters. The R-MDCT internal set included 512 3D stacks of ostiomeatal unit (OMU) MDCT head images acquired once from 512 participants. The external set consisted of 64 3D stacks from 64 subjects. The internal data included 254 maxillary sinus fungal balls (MFB) \cite{ferguson2000fungus}, 128 chronic rhinosinusitis (CRS), and 130 healthy control (HC) cases as disease classes, while the external data included 26 MFB, 18 CRS, and 20 HC cases. 

We requested annotations for the internal and external sets to ear, nose, and throat (ENT) physicians. Because each patient had a different disease status on the right or left side, each 3D OMU MDCT stack was divided into right and left halves, and each ENT physician was required to annotate either of the three disease classes on each side of the maxillary sinuses (i.e., HC, CRS, or MFB per side). Subsequently, individual annotations were confirmed by at least three experts for labeling after reaching a consensus. We obtained annotations from four ENT physicians with less than 10 years of experience and four specialists with more than 10 years of experience. In addition, for the AI-based CAD method to first distinguish slices corresponding to the maxillary sinuses before disease detection to constrain the disease area within the maxillary sinuses, we requested the ENT physicians to identify the entire (coronal view) slices in the R-MDCT samples showing the maxillary sinuses in both the internal and external sets.

\subsubsection{Data preprocessing: P-CBCT synthesis from R-MDCT}
\label{sec:preprocessing}

\begin{figure}[t]
\centering
\vspace{1cm}
\includegraphics[width=\linewidth]{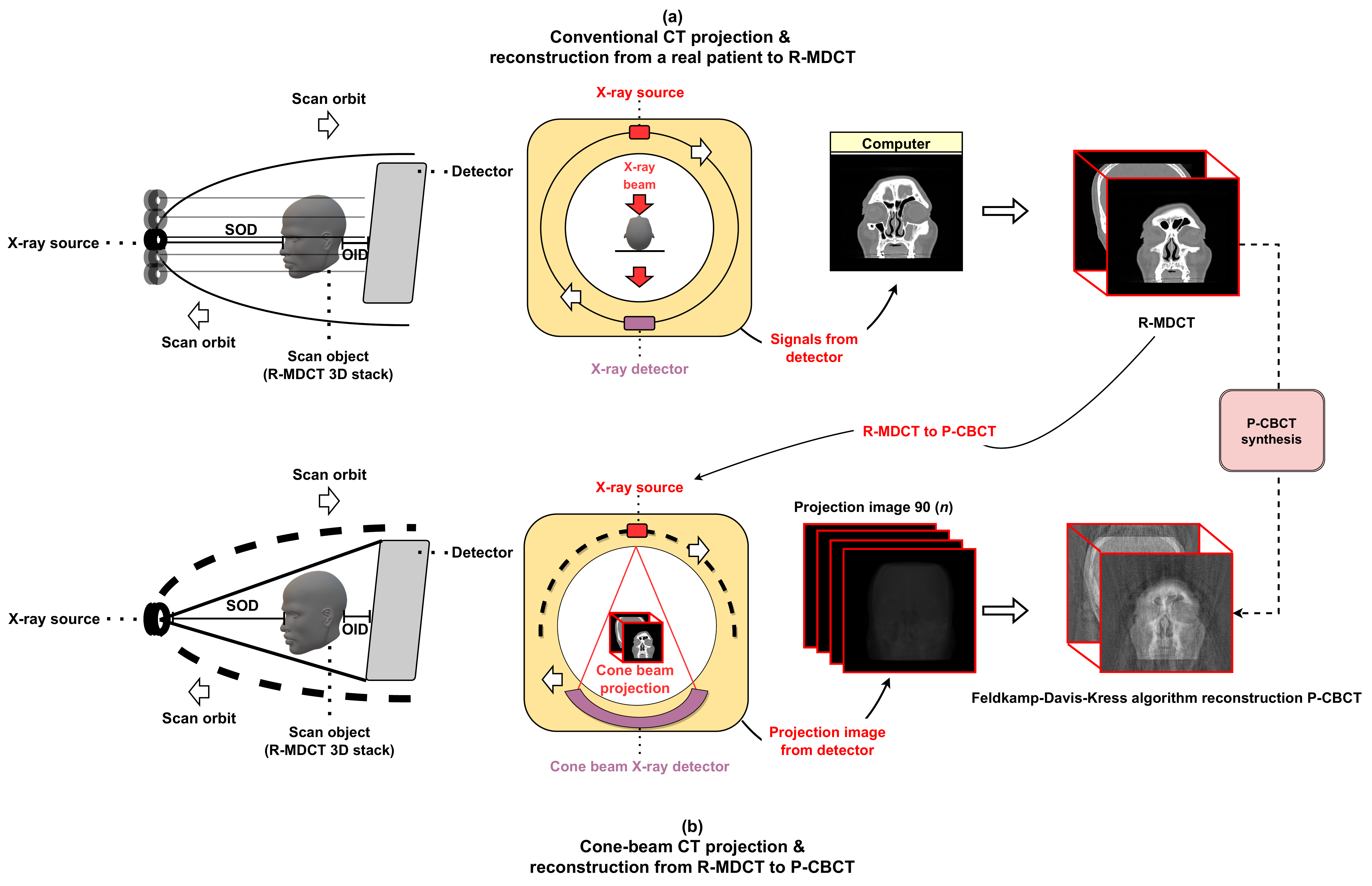} 
\caption{\textbf{P-CBCT image synthesized from R-MDCT scans based on 3D cone-beam projection and reconstruction.} (a) Schematic for obtaining R-MDCT data (conventional MDCT projection and reconstruction from an actual patient to R-MDCT). (b) Schematic for obtaining P-CBCT data (cone-beam MDCT projection and conversion from R-MDCT into P-CBCT data). (OID, object--image distance; SOD, source--object distance)}
\label{fig:CBCT_synthesis}
\end{figure}

As illustrated in Fig. \ref{fig:CBCT_synthesis}, we obtained the 3D stack of P-CBCT from the individual 3D stack of OMU MDCT data. To this end, each 3D stack of OMU MDCT was assumed to show the patient’s head, and CBCT was reconstructed using the projection images produced by executing a cone-beam projection. The projector rotated 180 degrees around the {cephalic} (i.e., R-MDCT 3D stack) about the horizontal axis, taking one shot every 2 degrees to collect 90 projection images before reconstruction using the Feldkamp--Davis--Kress \cite{scherl2007implementation} method to obtain a 3D stack of P-CBCT. For consistency with the actual imaging setup, we set the source--object distance (i.e., the distance between the X-ray source and patient), object--image distance (i.e., the distance between the patient and detector), and detector size to $1200$, $200$, and $512$~mm, respectively, using the ASTRA toolbox \cite{palenstijn2013astra}. As an obtained P-CBCT image was reconstructed from an existing R-MDCT image, the P-CBCT images used the annotations of the R-MDCT described in Section \ref{sec:data_ac_an}. Finally, the number of synthesized P-CBCT images was the number of acquired 3D stacks of R-MDCT. The internal P-CBCT set ($n$ = 512) was used for training, and the external P-CBCT set ($n$ = 64) was used for diagnostic performance evaluation.  
 
\subsection{AI-based method for lesion diagnosis on OMU CBCT}
\begin{figure}[t]
\centering
\includegraphics[width=\linewidth,height=6cm]{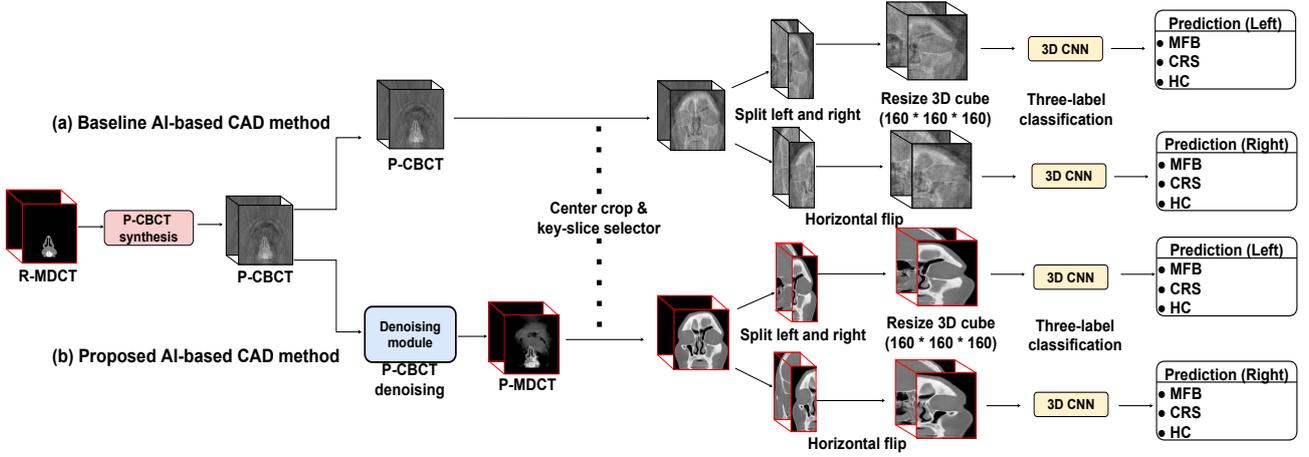} 
\caption{\textbf{Inference of (a) baseline and (b) proposed AI-based CAD methods.} The proposed method takes a 3D stack of OMU CBCT scans from a target subject as input. Unlike the baseline method, denoising is performed using a specific module (block in blue). A denoised 3D stack is derived as pseudo-MDCT. Using the 3D pseudo-MDCT stack, diagnosis is performed as in the baseline method \cite{kim2022detection}: extraction of the maxillary sinus region and diagnosis of diseases within that area through three-label classification.}
\label{fig:ai_inference}
\end{figure}

We introduce an advanced AI-based CAD method intended for CBCT and compare it with our previous development \cite{kim2022detection} considered as the baseline in the present study. Sections \ref{sec:inference} and \ref{sec:training} detail the diagnosis (inference) of the proposed method and its training processes, respectively, compared with those of the baseline. 

\subsubsection{Inference for CAD} 
\label{sec:inference} 
The inference processes of the baseline and proposed AI-based CAD methods are illustrated in Figs. \ref{fig:ai_inference}(a) and (b), respectively, and described below.

The baseline method receives a 3D OMU R-MDCT stack as input and generates a stack only containing the maxillary sinus region from a 2D convolutional neural network (CNN) \cite{chang2017efficient} that extracts only the slices with visible maxillary sinuses among the coronal slices (first stage). The diagnosis is predicted using a 3D CNN \cite{ren2018interleaved} that performs individual disease diagnosis for each sinus region (second stage).
Fig. \ref{fig:ai_inference}(a) describes this two-stage diagnosis process, which allows to perform accurate AI-based CAD of diseases within the region of interest (i.e., maxillary sinuses) in the original 3D OMU R-MDCT stack. The method achieves superior diagnostic performance to that of ENT resident physicians in a fully end-to-end approach, as the diagnosis can be done by directly using OMU MDCT images without any pretreatment assumptions.
The baseline can achieve accurate diagnosis without noise and artifacts, like in MDCT data, but it shows technical limitations for CBCT data that are generally polluted by strong noise and artifacts. For example, in some instances, the baseline shows poor diagnostic performance of lesions in some instances. In addition, the lesions are not apparent owing to CBCT noise, hindering the confirmation of diagnosis by a physician even when examining the region of interest on the CBCT image with the baseline results, indicating limited interpretive power. 

To improve the diagnostic performance and interpretive power, we incorporate denoising before diagnosis, establishing a novel AI-based CAD method that merges the two diagnosis stages preceded by the additional denoising stage for accurate diagnosis. The proposed method thus comprises the novel denoising (first stage) and the baseline stages of extracting the maxillary sinus region (second stage) and performing disease diagnosis in each maxillary sinus region (third stage). The processes of the proposed method are illustrated in Fig. \ref{fig:ai_inference}(b).

The proposed AI-based CAD method eliminates noise and artifacts from CBCT images, as illustrated in the first stage of Fig. \ref{fig:ai_inference}(b). Thus, it allows the synthesis of realistic pseudo-MDCT (P-MDCT) images. The P-MDCT images with mitigated noise promote disease identification by both a CAD method and a physician responsible for confirming the diagnosis. Therefore, the diagnostic performance of AI-based CAD and the physician's ability to interpret the results can be enhanced. This study is a pioneering demonstration of the importance of noise removal as an initial stage for AI-based CAD using CBCT, as verified by comparisons with results without denoising.

\subsubsection{Training}
\label{sec:training}

\begin{figure}[t]
\centering
\includegraphics[width=\linewidth, height=6cm]{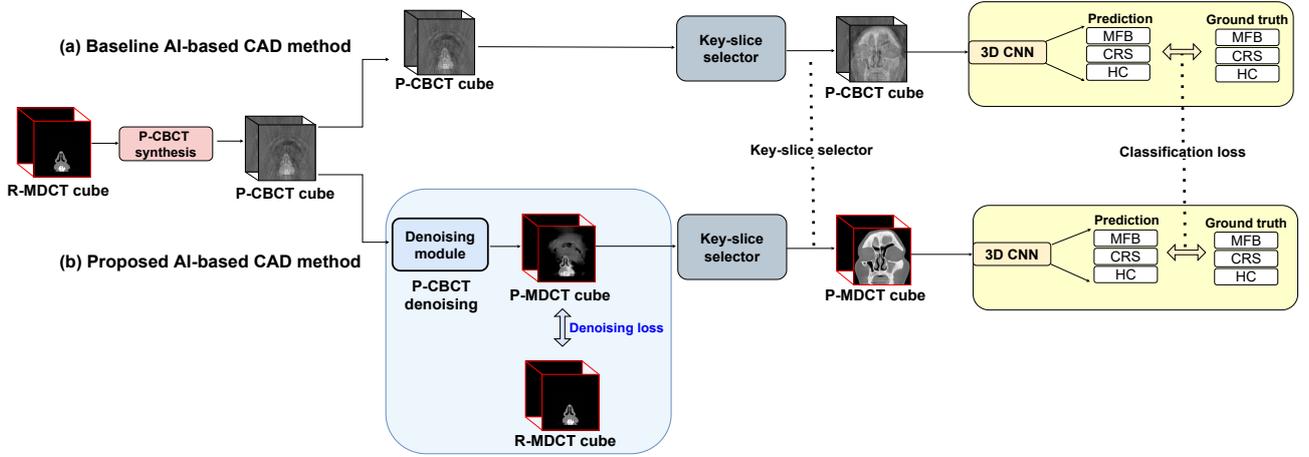}
\caption{\textbf{Training processes of (a) baseline and (b) proposed AI-based CAD methods.} Unlike the baseline, the proposed method includes a denoising module adding a denoising loss to the output. Both methods use a classification loss in the last stage for diagnosis.}  
\label{fig:training_process}
\end{figure}
We describe and compare the training processes of the proposed and baseline AI-based CAD methods. The processes are illustrated in Fig. \ref{fig:training_process}. Unlike the baseline, the proposed method includes training for denoising, while the remainder is the same as baseline training.

\paragraph{Denoising module.} To train the denoising module in the proposed method, we define a reconstruction loss such that the module output is an estimate of the 3D R-MDCT stack, called the 3D stack of pseudo-MDCT (P-MDCT), which represents the same data as the denoising module input, that is, the 3D P-CBCT stack. In other words, we generate the P-CBCT image from the R-MDCT image as described in Section \ref{sec:preprocessing} and then use the P-CBCT image as input of the denoising module to be trained aiming to obtain the original R-MDCT as its output. Training of the denoising module proceeds as follows: 
\begin{itemize}
    \item The denoising module is trained using multiscale U-Net-like sparse coding (MUSC) \cite{Liu2021learning}, which demonstrated the best performance compared with other denoising modules evaluated in this study. Nevertheless, other more advanced denoising models may be implemented in the future for the proposed method. Paired R-MDCT and P-CBCT images are used as training data because MUSC requires supervised learning. Training is performed in a 2D slice-by-slice process using the corresponding slices in each cube of R-MDCT and P-CBCT scans.
    \item We use the mean squared error in the denoising module as a representative reconstruction loss to generate P-MDCT from P-CBCT images. The denoising module accepts as input individual coronal 2D slices of each 3D stack resized to slices of $512 \times 512$ pixels and outputs same-sized slices. Training is conducted with a batch size of $18$, a learning rate of $10^{-4}$, $20$ epochs, and Adam optimization \cite{kingma2014adam}. We used the internal set with $512$ 3D stacks for training, and the external set for testing the trained denoising module.
\end{itemize}

\paragraph{Diagnosis module.} In the proposed method, the diagnosis module takes the noise-removed 3D stack from the denoising module as input and identifies the disease type (i.e., MFB, CRS, or HC) of the left and right maxillary sinuses. To predict the confirmed label provided by the physician (see Section \ref{sec:data_ac_an}) as the output of the diagnosis module, we assign a classification loss to the module output. The diagnosis module has a similar baseline to the proposed technique, but it receives the 3D CBCT stack as input, whereas the proposed technique receives the denoising output of this stack. Details of the diagnosis module can be found in our previous report \cite{kim2022detection}. The module consists of a key-slice selector followed by a multilabel classifier summarized as follows:

\begin{figure}[t]
\centering
\includegraphics[width=\linewidth, height=3.5cm]{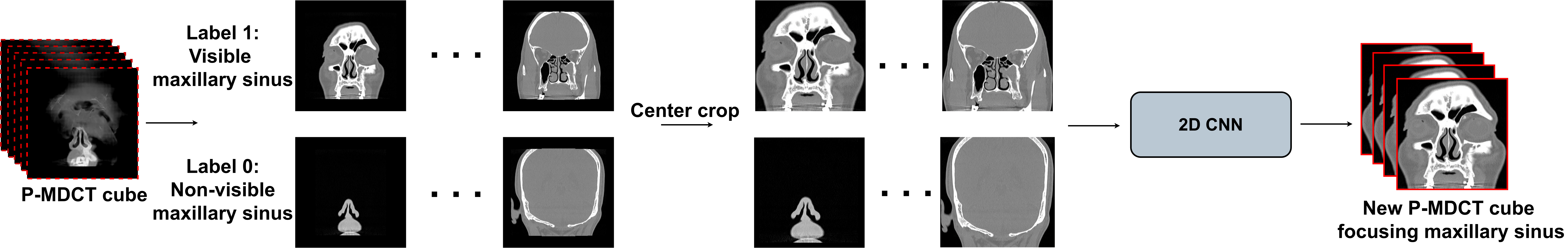}
\caption{\textbf{Process of key-slice selector.} This network extracts slices only including maxillary sinuses and generates the corresponding 3D P-MDCT stack. }
\label{fig:slice_selection}
\end{figure}

\begin{itemize}
    \item  Key-slice selector. The key-slice selector uses a 3D stack of head scans (i.e., 3D P-MDCT stack in the proposed method or 3D CBCT stack in the baseline method) from a patient as input. Only the coronal slices with the visible regions of maxillary sinuses are extracted to build the corresponding 3D stack. To this end, each coronal slice is used for training the key-slice selector to perform binary classification of the maxillary sinus presence. As a representative backbone network, we use Efficient-Net \cite{tan2019efficientnet}, which achieved the highest performance in our previous study \cite{kim2022detection}. Using the common classification loss given by cross-entropy \cite{de2005tutorial}, the model is trained to deliver label 1 if the maxillary sinus is visible and label 0 otherwise. The ground truth label is that provided by the physician. Fig. \ref{fig:slice_selection} illustrates key-slice selection, and additional training details are available in our previous study \cite{kim2022detection}. 
    \item  Multilabel classifier. We use a 3D stack centered on the maxillary sinuses obtained from the key-slice selector as input for the classification of another network, which predicts one of three labels for diagnosing the disease type in the left and right maxillary sinuses. As the delicate correlation between slices is important for the differentiation and detection of MFS and CRS, we use a 3D CNN instead of a 2D architecture, which was used in our previous study \cite{kim2022detection}. We divide the original 3D cube, including only the maxillary sinuses into two 3D cubes for the right and left sides, such that each maxillary sinus is centered, and the network receives the divided 3D cube as input for learning based on the cross-entropy loss to predict the disease type among MFB, CRS, and HC. We use 3D-ResNet  \cite{he2016deep} as the backbone of the 3D CNN using the weights from pretraining on the Kinetics dataset \cite{kay2017kinetics}. Additional training details are available in our previous study \cite{kim2022detection}. 
\end{itemize}

The training was conducted using fivefold cross-validation and the internal set,  and model evaluation was then performed on the external set.

\subsection{Evaluation metrics}
The following five statistical metrics were used to evaluate the performance of the classification model: accuracy, AUC of the receiver operating characteristic curve, sensitivity, precision, and F1 score. As the classification of three labels was performed, we used three groups for true positive (TP), false positive (FP), and false negative (FN) results with labels 1, 2, and 3 for MFB, CRS, and HC, respectively. One label was selected as positive and the other two as negative.  
\begin{gather}
Accuracy=\frac{\sum_{i=1}^C T_i}{D_{test}}\\
Precision=\frac{1}{C}\sum_{i=1}^C Precision_i=\frac{1}{C} \sum_{i=1}^C \frac{TP_i}{TP_i+FN_i}\\
Sensitivity=\frac{1}{C}\sum_{i=1}^C Sensitivity_i=\frac{1}{C}\sum_{i=1}^C \frac{TP_i}{TP_i+FN_i}\\
F1\_score=\frac{1}{C}\sum_{i=1}^C F1\_score_i=\frac{2}{C}\sum_{i=1}^C \frac{Precision_i \cdot  Sensitivity_i}{Precision_i+Sensitivity_i}
\end{gather}
where $C$ ($=3$) is the number of labels to be classified and $T_i$ is the number of correctly predicted samples for label $i$. Considering label $i$ as positive and the remaining labels as negative, $TP_i$, $FP_i$, and $FN_i$ represent the true positive, false positive, and false negative, respectively, and $Precision_i$, $Sensitivity_i$, and $F1\_score_i$ represent the precision, sensitivity, and F1 score for label $i$, respectively.

\section{Results}
\subsection{Quantitative evaluation}
We quantitatively compared the baseline and proposed AI-based CAD methods to determine whether the performance was enhanced by adding a denoising module. In addition, we compared the proposed CAD with the physician's diagnoses to evaluate the clinical usefulness of our proposal. These comparisons are respectively reported in Sections \ref{sec:q_model_comp} and \ref{sec:q_human_comp}.

\subsubsection{Diagnostic performance of baseline and proposed methods} 
\label{sec:q_model_comp}

\begin{table}[ht] 
\vskip -4pt
\caption{\textbf{Diagnostic performance of baseline and proposed AI-based CAD methods.} Micro- and macro-average AUCs obtained from external validation. Mean and standard deviation derived from fivefold cross-validation.} 
\footnotesize
\centering
{
	\resizebox{0.6\linewidth}{!}{
		\begin{tabular}{c|c}
			\toprule
			\bf{Metric} & \bf{External validation result} (\%) \\
            \midrule
			Micro-average AUC (Baseline) & $86.2\pm8.8$\\
            \hline
			Macro-average AUC (Baseline) & $87.0\pm4.1$\\
            \hline
            Accuracy (Baseline) & $73.4\pm9.34 $\\
            \hline
			Micro-average AUC (Proposed) & $\bf{93.6\pm1.7 \,(+7.4)}$\\
            \hline
			Macro-average AUC (Proposed) & $\bf{92.6\pm2.3 \,(+5.6)}$\\
            \hline
            Accuracy (Proposed) & $\bf{83.0\pm2.0 \,(+9.6)}$\\
            \hline
		\end{tabular}
  }
}
\label{tab:modelcomp_auc}
\end{table}

\begin{table}[htb!]
\vskip -4pt 
  \caption{\textbf{Diagnostic performance of baseline and proposed AI-based CAD methods.} Accuracy, sensitivity, precision, and F1 score obtained from external validation. Mean and standard deviation derived from fivefold cross-validation.}
  \centering
  \resizebox{0.6\linewidth}{!}{
  \begin{tabular}{c|c|c|c|c}
  \hline
  \hline
    \multicolumn{1}{c|}{\bf{Method}}
    & \multicolumn{4}{c}{\bf{Accuracy (\%)}}\\
    \hline
    \multicolumn{1}{c|}{Baseline} & \multicolumn{4}{c}{$73.4\pm10.4$}\\
    \hline
    \multicolumn{1}{c|}{Proposed} & \multicolumn{4}{c}{$\bf{83.0\pm2.3}$ $(9.6)\uparrow$}\\    
    \hline
    \hline
    \multicolumn{1}{c|}{\bf{Method}} &\multicolumn{4}{c}{\bf{Sensitivity} (\%) } \\
    \cline{2-5}
       & \bf{HC} & \bf{CRS} & \bf{MFB} & \bf{Average} \\\hline
       Baseline  & $83.0\pm20.8$ & $35.0\pm19.0$ & $69.3\pm21.8$ & $62.4\pm4.0$ \\
       \hline
       Proposed & $90.6\pm3.4$ & $69.5\pm13.3$ & $77.4\pm0.4$ & $\bf{79.2\pm3.6}$ $(16.8)\uparrow$\\
       \hline
    \multicolumn{1}{c}{} & \multicolumn{4}{c}{\bf{Precision} (\%)} \\
    \cline{1-5}
       Baseline & $91.9\pm4.1$ & $30.5\pm20.2$ & $73.2\pm15.9$ & $65.2\pm8.4$ \\
       \hline
       Proposed & $93.7\pm4.7$ & $66.7\pm5.8$ & $75.6\pm6.6$ & $\bf{78.7\pm2.8}$ $(13.5)\uparrow$ \\
       \hline
    \multicolumn{1}{c}{} & \multicolumn{4}{c}{\bf{F1 score } (\%)} \\
    \cline{1-5}
       Baseline & $85.8\pm12.9$ & $28.0\pm6.9$ & $67.4\pm12.6$ & $60.4\pm7.3$ \\
       \hline
       Proposed & $92.0\pm1.7$ & $67.3\pm6.8$ & $76.4\pm3.1$ & $\bf{78.6\pm3.1}$ $(18.2)\uparrow$ \\
       \hline
       \hline
  \end{tabular}
  }
  \label{tab:modelcomp_others}
\end{table}

\paragraph{Performance of baseline and proposed AI-based CAD methods.} 
The diagnostic performance results of the baseline and proposed AI-based CAD methods are listed in Tables \ref{tab:modelcomp_auc} and \ref{tab:modelcomp_others} regarding AUC and other metrics. As shown in Fig. \ref{fig:ai_inference}, the two methods use a 3D P-CBCT stack input and provide individual diagnostic outcomes for each sinus. We converted all the 3D R-MDCT stacks in the external stack into 3D P-CBCT stacks and used them as inputs. The average performance across sinuses was calculated for each of the five trained/cross-validation models, and their averages and standard deviations are also listed in the tables.

In Table \ref{tab:modelcomp_auc}, the proposed method increases the micro- and macro-average AUCs by 7.4\% and 5.6\%, respectively, compared with the baseline. In Table \ref{tab:modelcomp_others}, the proposed method enhances the performance by at least 10\% compared with the baseline for all the metrics of accuracy, specificity, sensitivity, and F1 score. These results verify the efficacy of the proposed method that adds the denoising module. 

\paragraph{Performance of different denoising models in the proposed method.}
\label{sec:Ablation_denoising}

To evaluate the backbone of the denoising module in the proposed method, we compared four backbone models: Cycle-GAN \cite{zhu2017unpaired}, DCL-GAN \cite{han2021dual}, U-Net \cite{gurrola2021residual},
CNCL-U-Net-denoising \cite{geng2021content}, and MUSC \cite{Liu2021learning}. In this evaluation, only annotated coronal 2D slices (paired R-MDCT and P-CBCT images) of the maxillary sinus were collected from the internal set. From the slices, 1000 were randomly selected and used for evaluation, and the rest were used for training. The two generative adversarial networks (GANs) are high-performance models in image translation that we used to convert a noisy image into a denoised image. We used the original GAN code and the associated loss in the evaluation. U-Net was trained as described in Section \ref{sec:training}. Table \ref{table:denoising} shows the denoising performance results (i.e., peak signal-to-noise ratio (PSNR) and structural similarity index (SSIM)) for each different model. The U-Net-based model showed higher performance than the GAN-based model, and MUSC showed the highest performance, which therefore was used in this study as a representative denoising model.


\begin{table}[t] 
\vskip -4pt
\caption{\textbf{Noise cancellation performance using different denoising models in the proposed technique.} PSNR \cite{huynh2008scope} and SSIM \cite{wang2003multiscale} were evaluated.}
\footnotesize
\centering
{
	\resizebox{0.5\linewidth}{!}{
		\begin{tabular}{c|c|c}
		    \hline
			\bf{Model} & \bf{PSNR} (dB) & \bf{SSIM} (\%)   \\
			\hline
            \multicolumn{3}{c}{ GAN-based } \\
            \hline
			Cycle-GAN \cite{zhu2017unpaired} & $12\pm0.9$ & $53\pm5.3$ \\
            \hline
            DCL-GAN \cite{han2021dual}  & $17\pm1.5$ & $58\pm5.7$ \\
            \hline
            \multicolumn{3}{c}{ U-Net-based } \\
            \hline
            U-Net \cite{gurrola2021residual} & $20\pm2.2$ & $81\pm5.0$ \\
            \hline
            CNCL-U-Net-denoising \cite{geng2021content} & $19\pm3.3$ & $83\pm6.6$\\
            \hline
			MUSC \cite{Liu2021learning} & $\bf{20\pm1.2}$ & $\bf{83\pm4.8}$\\
            \hline
		\end{tabular}
  }
}
\label{table:denoising}
\end{table}

\subsubsection{Diagnostic performance of proposed method and physician} 
\label{sec:q_human_comp}

\begin{table}[htb!]
\vskip -4pt 
  \caption{\textbf{Diagnostic performance of proposed AI-based CAD method and otorhinolaryngology resident physicians.} Micro- and macro-average AUCs obtained from external validation. Mean and standard deviation derived from fivefold cross-validation.}
  \centering
  \resizebox{0.6\linewidth}{!}{
  \begin{tabular}{c|c|c|c|c}
  \hline
  \hline
    \multicolumn{1}{c|}{\bf{Method}}
    & \multicolumn{4}{c}{\bf{Average accuracy (\%)}}\\
    \hline
    \multicolumn{1}{c|}{Manual} & \multicolumn{4}{c}{$71.7\pm2.8$ }\\
    \hline
    \multicolumn{1}{c|}{Proposed} & \multicolumn{4}{c}{$\bf{83.0\pm2.3}$ $(11.3)\uparrow$}\\    
    \hline
    \hline
    \multicolumn{1}{c|}{\bf{Method}} &\multicolumn{4}{c}{\bf{Sensitivity} (\%) } \\
    \cline{2-5}
       & \bf{HC} & \bf{CRS} & \bf{MFB} & \bf{Average} \\\hline
       Manual & $95.9\pm0.6$ & $52.5\pm14.8$ & $32.5\pm9.9$ & $60.3\pm3.4$ \\
       \hline
       Proposed & $90.6\pm3.4$ & $69.5\pm13.3$ & $77.4\pm0.4$ & $\bf{79.2\pm3.6}$ $(18.9)\uparrow$\\
       \hline
    \multicolumn{1}{c}{} & \multicolumn{4}{c}{\bf{Precision} (\%)} \\
    \cline{1-5}
       Manual & $94.3\pm0.0$ & $46.8\pm5.0$ & $40.0\pm8.8$ & $60.4\pm4.3$ \\
       \hline
       Proposed & $93.7\pm4.7$ & $66.7\pm5.8$ & $75.6\pm6.6$ & $\bf{78.7\pm2.8}$ $(18.3)\uparrow$ \\
       \hline
    \multicolumn{1}{c}{} & \multicolumn{4}{c}{\bf{F1 score } (\%)} \\
    \cline{1-5}
       Manual & $95.1\pm0.3$ & $49.1\pm9.1$ & $35.0\pm7.2$ & $59.7\pm3.1$ \\
       \hline
       Proposed & $92.0\pm1.7$ & $67.3\pm6.8$ & $76.4\pm3.1$ & $\bf{78.6\pm3.1}$ $(18.9)\uparrow$ \\
       \hline
       \hline
  \end{tabular}
  }
  \label{table:humanvs}
\end{table}

In addition to the comparative evaluation between the baseline and proposed methods, we compared the results from our proposal and external validation by six otolaryngology resident physicians to evaluate the clinical superiority of the proposed AI-based CAD method. The resident physicians who participated in this test were not involved in data extraction and annotation. In other words, the six physicians and the proposed method performed ternary classification of the disease type for each sinus across all 3D P-CBCT stacks in the external validation set.

Table \ref{table:humanvs} lists this evaluation's accuracy, specificity, and sensitivity results. The proposed method shows higher diagnostic performance for the three metrics than physicians by at least 10\%. In particular, it substantially improves the sensitivity and specificity by at least 18\%, demonstrating its clinical effectiveness. 

\begin{figure}[t]
\centering
\includegraphics[scale=0.17]{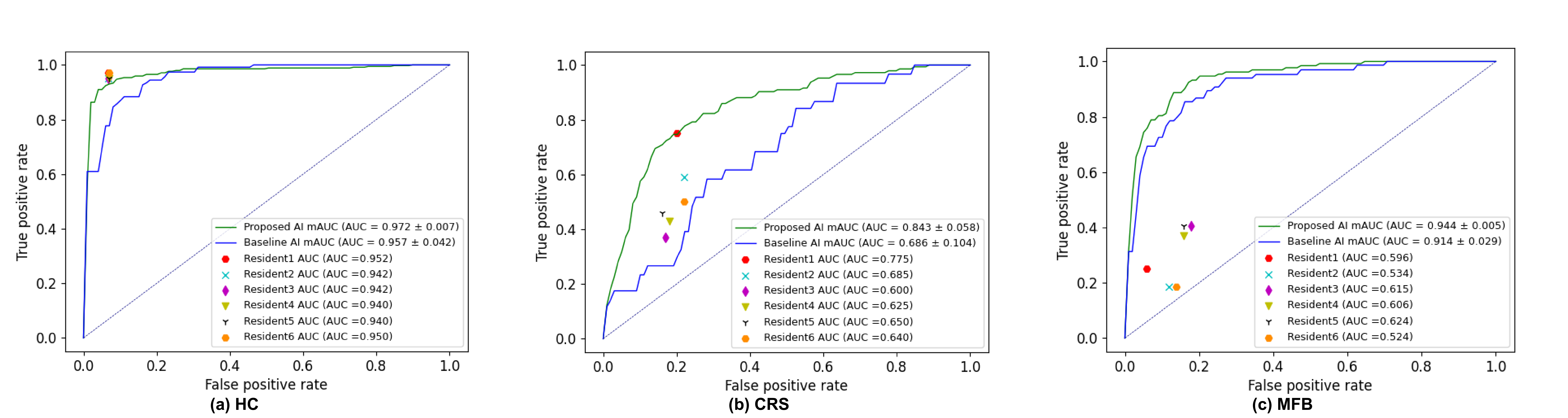}
\caption{\textbf{Diagnostic performance of proposed AI-based CAD method and six resident physicians regarding the receiver operating characteristic curve for each disease type.} The proposed method provides average AUCs (mAUCs) for (a) HC, (b) CRS, and (c) MFB of $0.972\pm0.007$ ($95$\% confidence interval, $0.958-0.986$), $0.843\pm0.058$ ($95$\% confidence interval, $0.729-0.957$), and $0.944\pm0.005$ ($95$\% confidence interval, $0.934-0.954$), respectively. }
\label{fig:auc_humanvs}
\end{figure}

In addition, the AUC was evaluated and compared for each disease class, obtaining the results shown in Fig. \ref{fig:auc_humanvs}. The proposed method demonstrates comparable sensitivity/specificity in the normal group to the physicians and superior sensitivity/specificity for disease detection. In particular, for CRS detection, the baseline achieves a lower detection than a resident physician. In contrast, the proposed method has an average detection rate of at least 15\% higher than that of the resident physician. These results confirm that the proposed method can outperform both human diagnosis and the existing CAD method.

\subsection{Qualitative evaluation}
\label{sec:qual}
We also performed qualitative comparisons for various coronal scans.

\subsubsection{Gradient-weighted class activation maps of baseline and proposed methods}

\begin{figure}[ht]
\centering
\includegraphics[width=.9\linewidth]{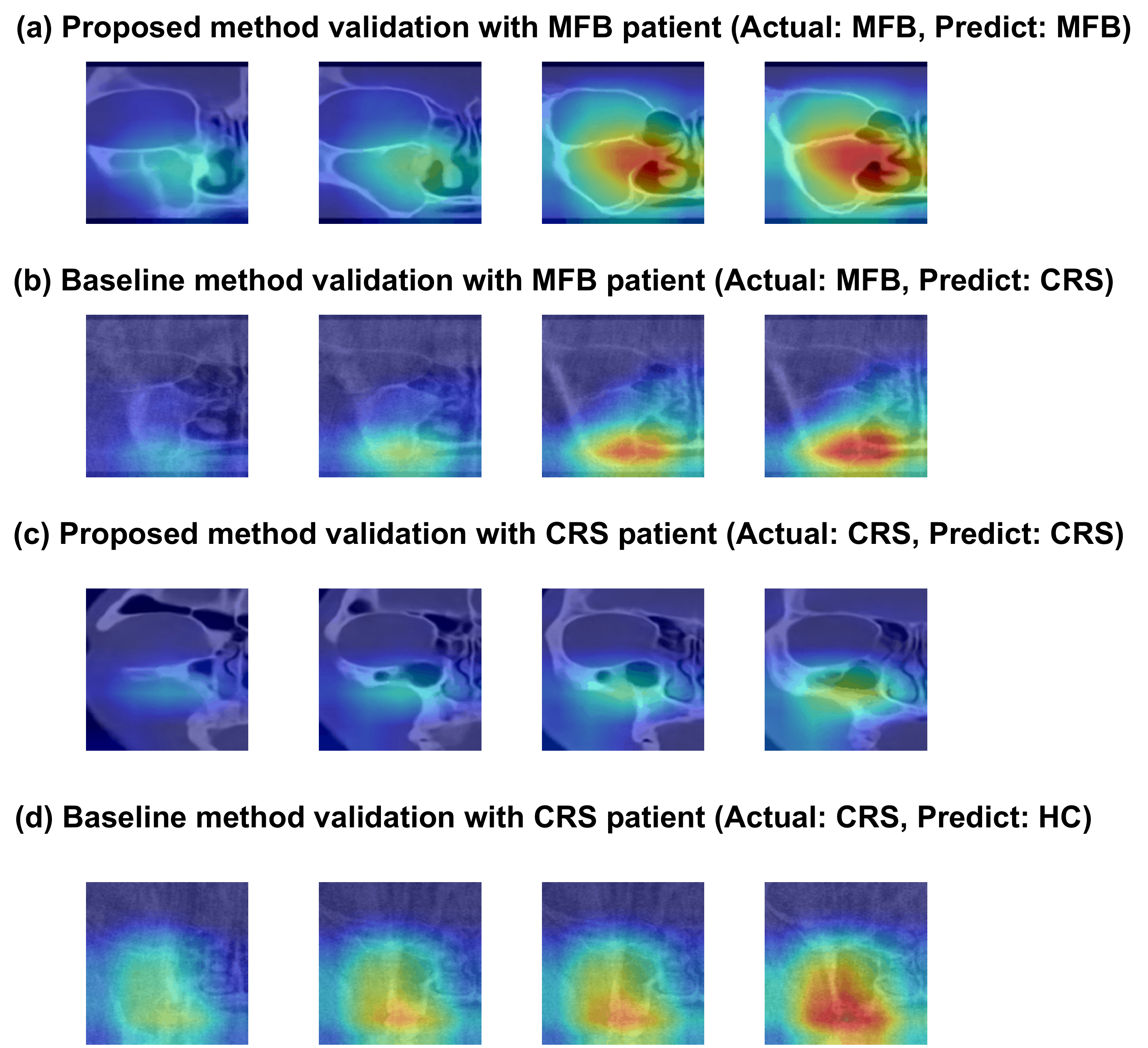} 
\caption{\textbf{Activation maps of diagnoses obtained from baseline and proposed methods.} Activation maps for (a,c) proposed and (b,d) baseline methods.}
\label{fig:gradcam}
\end{figure}

Fig. \ref{fig:gradcam} shows activation maps obtained from the baseline and proposed AI-based CAD methods. The maps show relevant parts of the image for AI-based diagnosis obtained by gradient-weighted class activation mapping, as in our previous study \cite{kim2022detection}. The proposed method correctly detects diseased maxillary sinuses, as highlighted in the maps, and correctly classifies the disease type (i.e., CRS or MFB). On the other hand, the baseline method fails to detect the maxillary sinus and classify the disease.

Unlike the baseline, the proposed method provides denoised scans (P-MDCT images) as input CBCT scans. Fig. \ref{fig:gradcam} shows that the proposed method uses denoised scans as background in each activation map, whereas the baseline method should rely on the original CBCT scan as background. Hence, the activation maps obtained from the proposed method show areas with apparent diseases on the denoised image. When a physician makes a final diagnosis by referring to the CAD results, the proposed method allows the examination of denoised scan findings in highly active map areas, thus facilitating the CAD interpretation by the physician. In contrast, the baseline method provides activation maps considering raw CBCT scans that inevitably contain noise and artifacts. As noise and distortions pollute the activation map, the physician cannot easily assess the result accuracy and interpret the diagnosis.

\subsubsection{P-MDCT image estimated by denoising R-CBCT image}
The proposed AI-based CAD method improves the diagnostic performance on CBCT scans compared with the baseline method and manual assessment. Hence, P-MDCT image estimation from a P-CBCT scan is validated. Nevertheless, the P-MDCT estimate from the denoising module in the proposed method is inferred from a P-CBCT image (i.e., CBCT scans are synthesized from MDCT scans). To evaluate this process, we tested noise removal of the trained denoising module in the proposed method for R-CBCT images in addition to P-CBCT images.  

\begin{figure}[t]
\centering
\includegraphics[width=.7\linewidth]{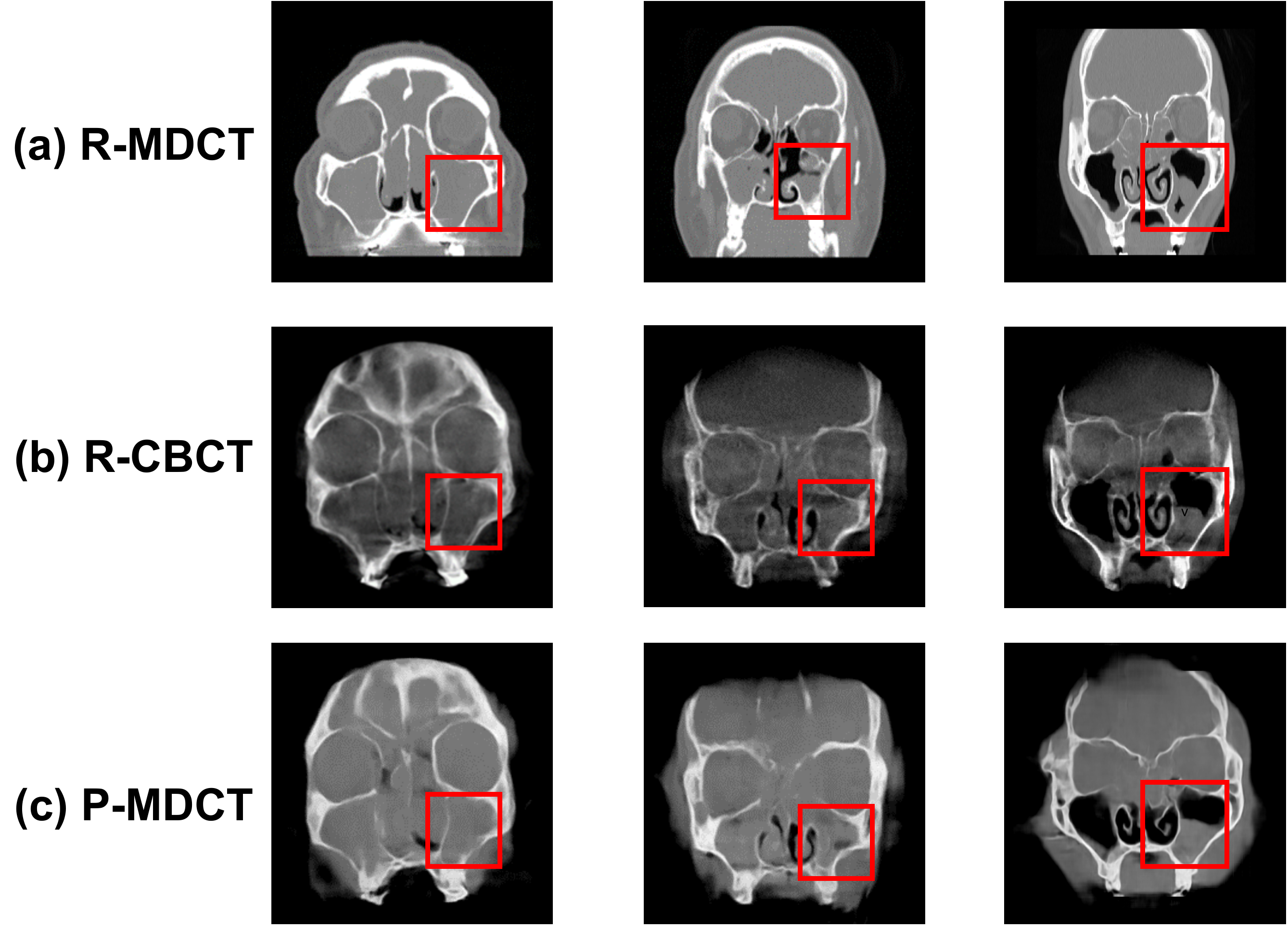} 
\caption{\textbf{Denoising examples of R-CBCT images.} (a) R-MDCT, (b) R-CBCT, and (c) P-MDCT (denoised R-CBCT) images.}
\label{fig:comparison_realworld}
\end{figure}

For this evaluation, physicians at our institution acquired three sets of patient-matched 3D stacks of R-MDCT and R-CBCT. The denoising module used the training weights obtained for diagnosis evaluation (see Section \ref{sec:qual}). Thus, R-CBCT data were not used for training the denoising module. Figs. \ref{fig:comparison_realworld}(a)--(c) show R-MDCT, R-CBCT, and denoised R-CBCT images (denoted as P-MDCT images because they are the output of the trained denoising module), respectively. In P-MDCT images, noise in R-CBCT images is removed, and the same anomaly structure of the R-MDCT images is observed (red boxes). Thus, the denoising module can take an R-CBCT image as input and perform denoising to determine the corresponding P-MDCT image with no substantial disease-related artifacts. Consequently, P-MDCT images can be obtained not only using P-CBCT images but also R-CBCT images, demonstrating the wide applicability of the denoising module.
 
\section{Discussion}

\paragraph{Clinical implications.}
This study is the first to improve the discrimination of soft tissues, which is difficult in CBCT, through an AI-based method and improve CAD of lesions in the maxillary sinus, providing great clinical significance. As the demand for accurate diagnosis of CRS increases, the number of MDCT examinations is also increasing. However, CBCT is a more convenient alternative that uses less radiation at the expense of lower resolution around soft tissues and signal-to-noise ratio than MDCT. In addition, a fungal ball in the maxillary sinus is a disease that does not respond to drug treatments, including antibiotics, and thus requires surgery \cite{fadda2021treatment}. Therefore, early diagnosis is essential but often delayed given the difficulty of accurate diagnosis in CBCT images \cite{hodez2011cone}. The proposed AI-based CAD method may be implemented in CBCT equipment and possibly support clinicians in the accurate identification of diseases that require surgery in the early stages.
  
\paragraph{Technical implications.}
This study has demonstrated the feasibility of using CBCT to detect maxillary sinus diseases by mitigating image noise, increasing the accuracy by 11.3\% and 9.6\% compared with manual assessment and a baseline method, respectively. Nevertheless, as the experimental evaluation only considered P-CBCT data (i.e., CBCT images synthesized from R-MDCT data) and not R-CBCT data, its validation with actual clinical data remains to be conducted.

\begin{figure}[t]
\centering
\includegraphics[width=\linewidth]{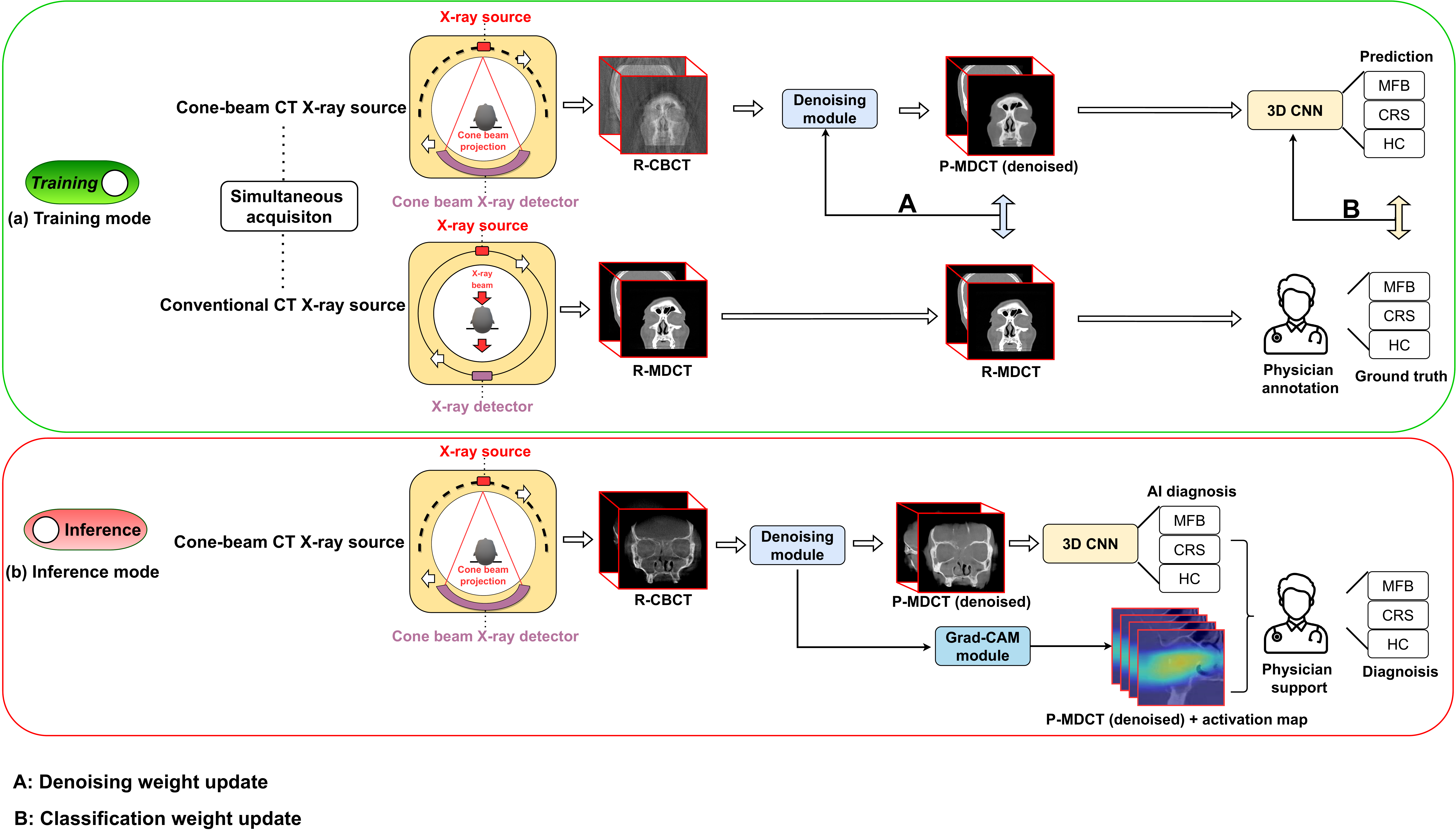} 
\caption{\textbf{New approach for radiographic image reconstruction using proposed AI-based CAD method}. (a) Simultaneous imaging of CBCT and MDCT for training diagnosis method. (b) Simultaneous acquisition of virtual MDCT and diagnoses from trained model using only CBCT image as input. (Grad-CAM, gradient-weighted class activation mapping)}
\label{fig:guide}
\end{figure}

Despite current limitations, we provide a new approach for developing CAD in medical devices, as illustrated in Fig. \ref{fig:guide}. If radiology medical equipment can acquire paired R-CBCT and R-MDCT scans, P-CBCT data used in this study can be substituted with R-CBCT data. Hence, if the proposed AI-based CAD method is applied to such a medical device, the denoising module can be trained to synthesize R-MDCT images from R-CBCT scans, and the subsequent diagnosis module can be trained to infer the corresponding diagnosis using the R-MDCT estimate as input. Therefore, a device capable of simultaneous CBCT and MDCT enables training of the CAD method using R-CBCT data (and its corresponding R-MDCT data) instead of P-CBCT data. The trained model performs diagnosis directly using R-CBCT images instead of R-MDCT images, allowing diagnosis at a low radiation dose. 

In its current form, the proposed method provides not only a diagnosis result but also a denoised P-MDCT image and activation map from an R-CBCT image. Thus, practitioners can receive valuable assistance for interpreting AI CAD even from a CBCT scan. The benefits of the proposed method may justify the corresponding modification of medical device usage.

\section{Conclusion}
We demonstrate that eliminating noise from CBCT images can enhance the performance of AI-based CAD with a model-agnostic method. The proposed method justifies and validates the inclusion of noise removal for diagnosis in soft tissues using CBCT. Our results suggest a successful diagnosis of soft tissue lesions of the sinonasal area, outperforming a baseline method and specialists. We expect that the method will enable accurate diagnosis using only low-dose CBCT, thereby reducing the burden of examination for both clinicians and patients, particularly in locations with shortages of physicians and high-dose screening systems.

\section*{Acknowledgements}
This work was supported by a grant from the National Research Foundation of Korea (NRF) funded by the Korean government (MSIT) (2021R1F1A106153511), by a grant from Korea Medical Device Development Fund funded by the Korean government (Ministry of Science and ICT, Ministry of Trade, Industry and Energy, Ministry of Health $\&$ Welfare, Ministry of Food and Drug Safety) (202011B08-02, KMDF$\_$PR$\_$20200901$\_$0014-2021-02, RS-2020-KD000014), by the Technology Innovation Program (20014111) funded by the Ministry of Trade, Industry $\&$ Energy (MOTIE, Korea), and by the Future Medicine 20*30 Project of Samsung Medical Center (SMX1210791).

\section*{Data availability}
The authors declare that the main data supporting the results of this study are available within the paper. The raw datasets are not widely available due to patient confidentiality and hospital privacy policy. The authors have no right to provide material to another person or institution.

\section*{Code availability}
The code used for the DL models is available on GitHub (To be disclosed upon submission of the revised version): \href{https://github.com/kskim-phd/CBCT-DENOISING-CAD}{\texttt{CBCT-DENOISING-CAD}}.

\clearpage 
\bibliographystyle{unsrtnat}
\bibliography{refs.bib}
\clearpage 

\appendix
	
\end{document}